\documentclass[%
 aip,
 jmp,%
 amsmath,amssymb,
preprint,%
]{revtex4-1}

\usepackage{graphicx}
\usepackage{dcolumn}
\usepackage{bm}
\usepackage{epstopdf}
\usepackage{amssymb}
\usepackage{color}
\usepackage[colorlinks=true, letterpaper=true, pdfstartview=FitV, linkcolor=blue, citecolor=blue, urlcolor=blue]{hyperref}

\begin{document}

\title{Electronic, magnetic, and optical properties of Mn-doped GaSb: a first-principles study}

\author{Chuang Wang}
\author{Wenhui Wan}
\author{Yanfeng Ge}
\affiliation{State Key Laboratory of Metastable Materials Science and Technology \& Key Laboratory for Microstructural Material Physics of Hebei Province, School of Science, Yanshan University, Qinhuangdao 066004, China }
\author{Yong-Hong Zhao}
\affiliation{College of Physics and Electronic Engineering, Center for Computational Sciences, Sichuan Normal University, Chengdu, 610068, China}
\author{Kaicheng Zhang}
\affiliation{College of Mathematics and Physics, Bohai University, Jinzhou 121013, China}
\author{Yong Liu}\email{yongliu@ysu.edu.cn, or ycliu@ysu.edu.cn}
\affiliation{State Key Laboratory of Metastable Materials Science and Technology \& Key Laboratory for Microstructural Material Physics of Hebei Province, School of Science, Yanshan University, Qinhuangdao 066004, China }

\begin{abstract}
 Half-metallic ferromagnets can produce fully spin-polarized conduction electrons and can be applied to fabricate spintronic devices. Thus, in this study, the electronic structure, magnetic properties, and optical properties of GaSb, which has exhibited half-metallicity, doped with Mn, a 3d transition metal, are calculated using the generalized gradient approximation and Heyd-Scuseria-Ernzerhof (HSE) functional. Ga$_{1-x}$Mn$_x$Sb ($x = 0.25, 0.5, 0.75$) materials exhibit ferromagnetic half-metallic properties and a high Curie temperature, indicating that this series can applied in spintronic devices. Meanwhile, they absorb strongly in the infrared band, suggesting that Ga$_{1-x}$Mn$_{x}$Sb also has potential applications in infrared photoelectric devices.
\end{abstract}


\maketitle


\maketitle

\section*{Introduction}

Over the past few decades, spintronics has developed rapidly. Compared with traditional semiconductor devices, spintronic devices have the advantages of storage non-volatility, lower power consumption, higher integration, etc.~\cite{1,2}. In 1983, through their study of alloys such as Heusler alloys NiMnSb and PtMnSb, Groot et al. first discovered a material with a special energy band structure in which one direction of the electron spin band exhibits metallicity, while the other direction of the electron spin band is semiconducting. They thus named the material a ¡°half-metallic ferromagnet (HMF)¡± ~\cite{3,4}. This type of material can produce fully spin-polarized conduction electrons, a good source of spin-flow injection~\cite{5,6,7}, With their magnetic moment quantization and zero magnetic susceptibility, HMFs can be used to fabricate spintronic devices such as spin diodes, spin field effect transistors, spin valves, and spin filters~\cite{8,9,10,11}. Thus, not only will HMFs play an important role in a new generation of high-performance microelectronic devices but they will also open up new avenues in the research of polarization transport theory and spintronics~\cite{12,13,14,15}.

Unlike conventional electronics, spintronics utilizes the charge and spin of electrons to carry information and, thus, has superior performance. During the development of spintronic materials, new magnetic materials with both magnetic and semiconductor properties have been discovered, which has aroused great interest in the study of spintronics. Binary semiconductors doped with magnetic transition metal elements have also played an important role in spintronics development. Related research has found that a small number of magnetic elements can be incorporated into semiconductors such as Group II-VI, Group IV, or Group III-V elements~\cite{16,17,18}. In such systems, the incorporated magnetic atoms replace the cations or anions in the semiconductor unit cell, or defects form in the system, both of which have led to the discovery of many new spintronic materials. Investigations have revealed that III-V compound semiconductor materials have broad applications in optoelectronic devices, optoelectronic integration, ultrahigh-speed microelectronic devices and ultrahigh-frequency microwave devices and circuits ~\cite{19,20}.

In recent years, many new half-metallic (HM) materials have been discovered by doping III-V binary semiconductors with transition metal elements~\cite{21,22,23,24}. However, few studies exist in the literature on the electronic properties of the III-V semiconductor material GaSb doped with 3d transition metals. GaSb is a direct band gap semiconductor with a zinc-blende (ZB) crystal structure, a band gap of about 0.72 eV, and a lattice constant of 0.61 nm~\cite{25}, GaSb has the characteristics of high electron mobility, high frequency and low threshold, and a high photoelectric conversion rate ~\cite{26,27,28}. Meanwhile, the lattice constant of this material approximately matches those of other various ternary and quaternary III-V compound semiconductor materials, which can greatly mitigate the problematic stress and defects caused by lattice mismatch. Thus, GaSb has become an important substrate material for preparing long-wave light-emitting diodes (LEDs), photodetectors, and fiber-optic communication devices~\cite{29,30,31}. Nonetheless, to the best of our knowledge, GaSb doped with the transition metal Mn has never been systematically researched about~\cite{32,33,34,35}. In this paper, we report first-principles calculations of Ga$_{1-x}$Mn$_{x}$Sb ($x = 0.25, 0.5, 0.75$) compounds, which are observed to be HMFs and, thus, could be useful in both spintronics and infrared photoelectrics.

The remaining part of the paper is organized as follows: in Section II, we present our calculations in detail; in Section III, we describe the results and discuss the electronic, magnetic, and optical properties of these ZB ternary Ga$_{1-x}$Mn$_{x}$Sb compounds. Finally, in Section IV, we summarize our results.

\section*{Methods}
GaSb has the ZB crystal structure, and its space group is $F \bar{4} 3m$ (No. 216)~\cite{36, 37}. Each Ga (Sb) atom is at the center of a tetrahedron of its four nearest-neighbor Sb (Ga) atoms, as shown in Fig.1. For the calculations, we replace either one, two, or three Ga atoms with Mn atoms in the GaSb unit cell, for a Mn doping ratio of $25\%$, $50\%$, and $75\%$, respectively. Thus, we obtain a series of ZB Ga$_{1-x}$Mn$_{x}$Sb ($x = 0.25, 0.5, 0.75$) compounds. The main contents of this paper are the study of the electrical, magnetic, and optical properties of ZB Ga$_{1-x}$Mn$_x$Sb ($x = 0.25, 0.5, 0.75$) by first principles.

\begin{figure}[htb]
  \centering
  \includegraphics[width=.35\textwidth]{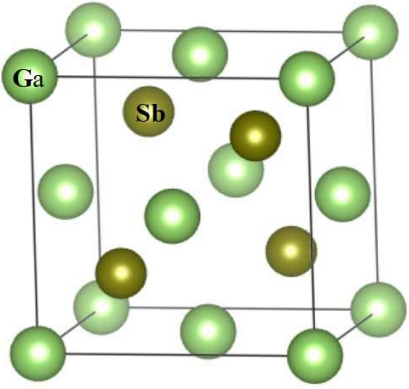}%
 \caption{Structure of zinc-blende GaSb.}\label{fig1}
\end{figure}

All calculations were performed using the first-principles calculation package VASP (Vienna ab initio simulation package)~\cite{38,39}. For the simulation, we chose the projected augmented wave (PAW)~\cite{40} to describe the interaction between electrons and nuclei. The correlation between electrons and electrons uses a generalized gradient (generalized gradient approximation, GGA)~\cite{41} in the form of the Perdew-Burke-Ernzerhof functional (PBE-96)~\cite{42}. First, we optimize the Monkhorst-Pack special K-point selection for the plane wave truncation energy and the Brillouin zone integral for each structural system. E$_{cut}$ is set to 400 eV, and the K point is $7 \times7 \times 7$. In the crystal structure optimization and atomic relaxation process, the accuracies of the total energy convergence of the system and the force convergence on a single atom are $10^{-5}$ eV and $10^{-4}$ eV/atom, respectively.

We calculated the total energy of non-magnetic (NM), ferromagnetic (FM), and antiferromagnetic (AFM) states of ZB Ga$_{1-x}$Mn$_{x}$Sb ($x = 0.25, 0.5, 0.75$) as a function of the lattice constant. Because the PBE functional calculation underestimates the crystal band gap, we added the Heyd-Scuseria-Ernzerhof (HSE) hybrid density functional~\cite{43,44} to calculate the electron band. HSE is used to solve the numerical analysis of the Kohn-Sham~\cite{45} equation in the plane wave basis set and introduces the non-local exact exchange. Some of this equation is calculated by a precise exchange, thus, improving the electron self-interaction of the system. The total energy of the electronic system can be described well by HSE, and the calculated energy band is similar to that observed experimentally. Thus, the calculation results are reasonable.

\section*{Results and Discussions}

Magnetic properties.
We calculated the electronic ground state properties of Ga$_{1-x}$Mn$_{x}$Sb materials at different Mn concentrations. As Fig. 2(d) shows, the FM and AFM states are calculated mainly by making the spins of the even-numbered Mn parallel, anti-parallel, and cross-parallel. Figure 2(a), 2(b) and 2(c) show the energy¨Cvolume curves corresponding to the three different magnetic states (NM, FM, AFM, respectively) of the Ga$_{1-x}$Mn$_{x}$Sb at different Mn concentrations. The figure demonstrates that the FM state has the lowest energy in all three states. Therefore, for equilibrium lattice constant, the FM state of ZB Ga$_{1-x}$Mn$_{x}$Sb is the most stable.

\begin{table}[h] 
\newcommand{\PreserveBackslash}[1]{\let\temp=\\#1\let\\=\temp}
\newcolumntype{C}[1]{>{\PreserveBackslash\centering}p{#1}}
\newcolumntype{L}[1]{>{\PreserveBackslash\raggedright}p{#1}}
\newcommand{\tabincell}[2]{\begin{tabular}{@{}#1@{}}#2\end{tabular}} 
\caption{ Ga$_{1-x}$Mn$_{x}$Sb ($x = 0.25, 0.5, 0.75, 1$) magnetic moment M$_{tot}$/N$_{Mn}$, Mn atom d-orbit magnetic moment M$_{Mn}$, Sb atom p-orbit magnetic moment M$_{Sb}$, ground state magnetic properties (GP) and Curie temperature T$_{C}$ (K). Material properties (MPs) are also given. }\label{table I}
\begin{ruledtabular}
\begin{tabular}{lcccccccccccc}
x & M$_{tot}$/N$_{Mn}$ ($\mu_{B}$) & M$_{Mn}$ ($\mu_{B}$) & M$_{Sb}$ ($\mu_{B}$)& GP & T$_{C}$ & MP \\
\hline
0.25 & 4 & 3.831 & -0.075 & FM & 496 & HMF\\
0.5  & 4 & 3.918 & -0.124 & FM & 866 & HMF \\
0.75 &4 & 3.889 & -0.156 &  FM & 919 & HMF \\
1    & 4~\cite{46} & - & - & FM & 587~\cite{47} & HMF \\
\hline
\end{tabular}
\end{ruledtabular}
\end{table}

\begin{figure}[htb]

  \centering
  \includegraphics[width=.45\textwidth]{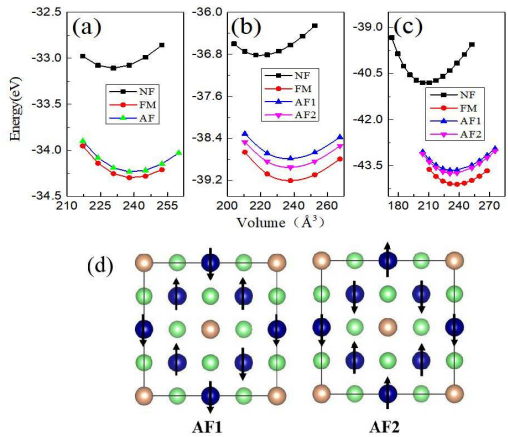}
 \caption{Energy-volume curves of  Ga$_{1-x}$Mn$_x$Sb ($x = 0.25, 0.5, 0.75$) (a) Ga$_{0.75}$Mn$_{0.25}$Sb; (b) Ga$_{0.5}$Mn$_{0.5}$Sb; (c) Ga$_{0.25}$Mn$_{0.75}$Sb; (d) AF1 and AF2 stand for two different AFM states.}\label{fig2}
\end{figure}

In their equilibrium lattice constants, the Ga$_{1-x}$Mn$_{x}$Sb ($x = 0.25, 0.5, 0.75$) compounds have an integer magnetic moment (in a Bohr magnet, $\mu_{B}$), corresponding to the FM HM nature. The average magnetic moment of a single Mn atom in Ga$_{1-x}$Mn$_{x}$Sb ($x = 0.25, 0.5, 0.75$) is 4.0 $\mu_{B}$. Fig.3 shows the total magnetic moment of these three FM HM materials, which varies with the lattice constant within $¡À20\%$. The total magnetic moment remained constant for lattice changes between $-5\%$ and $25\%$. Additionally, we explored the magnetic moment contribution of each atom in the material. The total magnetic moment is dominated by the magnetic moment of the Mn-$d$ orbital, and the contributions of other atomic magnetic moments are small. Furthermore, the contributions of the Mn-$d$ and Sb-$p$ orbital magnetic moments are opposite each other. Fig.3 demonstrates that the magnetic moment changes of Mn-$d$ increase with the increasing lattice constant, but this trend is gradual. Table I lists the magnetic moments and material properties of each atom. Next, we estimate the Curie temperature of the three FM half-metals using the mean field method~\cite{48,49,50},

\begin{eqnarray}
\frac{3}{2}k_{B}T&=& -\frac{\bigtriangleup E_{FM-AF}}{C} \label{equ1}
\end{eqnarray}

where C represents the number of doped Mn atoms in the unit cell, and $k_{B}$ represents the Boltzmann constant. All three compounds are found to have a Curie temperature above 400 K at room temperature. MnSb has been proven to be a FM half-metal with a Curie temperature of 587 K~\cite{47}. Table I shows the magnetic properties and Curie temperature of Ga$_{1-x}$Mn$_x$Sb ($x = 0.25, 0.5, 0.75, 1$).

Electronic properties.
Fig.4 gives the energy band structures of Ga$_{1-x}$Mn$_{x}$Sb ($x = 0.25, 0.5, 0.75$) compounds at their equilibrium lattice constants. The spin-up energy bands of Ga$_{1-x}$Mn$_{x}$Sb ($x = 0.25, 0.5, 0.75$) straddle the Fermi level E$_{F}$, which is metallic, while the spin-down energy band has a gap, which indicates insulator behavior. Ga$_{1-x}$Mn$_{x}$Sb has a conduction band bottom and a valence band top located at the high symmetry point $\Gamma$ of the Brillouin zone, which is a direct band gap, and the band gap becomes wider as the Mn concentration increases, as shown in Table I. Therefore, the different characteristics exhibited by the spin-up and spin-down electron energy bands reveal that the ZB Ga$_{1-x}$Mn$_{x}$Sb crystal has HM properties. We also used the same method to calculate the electronic structure of ZB GaSb and ZB MnSb by completely replacing Ga atoms, and we obtained the band diagram of the semiconductor GaSb with the direct band gap at the center of the Brillouin zone. MnSb is a FM half-metal, as shown in Fig.4(d), which agree with previous results. These two crystalline materials have long been studied, and our results are consistent with the electronic structures reported in the previous literature~\cite{37,46}. By comparison, we found that the FM HM properties of Ga$_{1-x}$Mn$_{x}$Sb ($x = 0.25, 0.5, 0.75, 1$) are caused by Mn atoms.

\begin{figure}[htb]
  \centering
  \includegraphics[width=.5\textwidth]{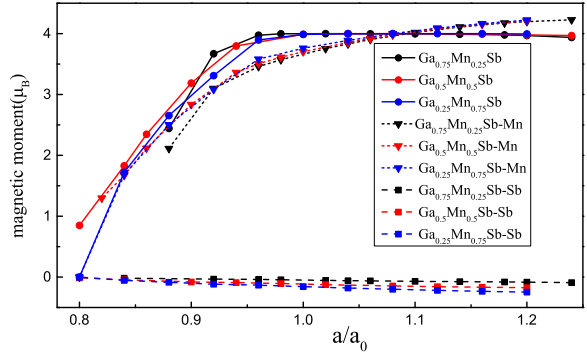}
  \caption{Total magnetic moment and the contribution of the magnetic moment from the Mn-$d$ and Sb-$p$ orbits as a function of the relative change in the lattice constant.}\label{fig3}
\end{figure}

Table II shows the equilibrium lattice constants and ground state properties calculated for Ga$_{1-x}$Mn$_{x}$Sb ($x = 0, 0.25, 0.5, 0.75, 1$) and related values from the literature.
We determined that the Ga-Sb bond is 2.638 $\AA$ long when the semiconductor GaSb is for the equilibrium lattice constant. By comparing the bond lengths among the atoms in the series of FM Ga$_{1-x}$Mn$_{x}$Sb ($x = 0.25, 0.5, 0.75, 1$) shown in Table II, we can see that after doping with Mn, the outermost orbital electrons of Mn in Ga$_{1-x}$Mn$_{x}$Sb ($x = 0.25, 0.5, 0.75, 1$) are consumed by the bond owing to the difference in electronegativity and the electron orbital interatomic hybridization. Because a larger Ga atom is substituted with a smaller Mn atom, the Mn-Sb bonds in the ZB structure are shorter than the Ga-Sb bonds. Table II demonstrates that the semiconductor band gap and the HM gap of HMF Ga$_{1-x}$Mn$_{x}$Sb ($x = 0.25, 0.5, 0.75, 1$) both increase with increasing concentration. The minimum of the HM gap of a HM material is intermediate along the distance from the Fermi energy to either the valence band or the bottom of conduction band in the spin sub-band with a band gap~\cite{51}.

\begin{figure}[htb]
  \centering
  \includegraphics[width=.4\textwidth]{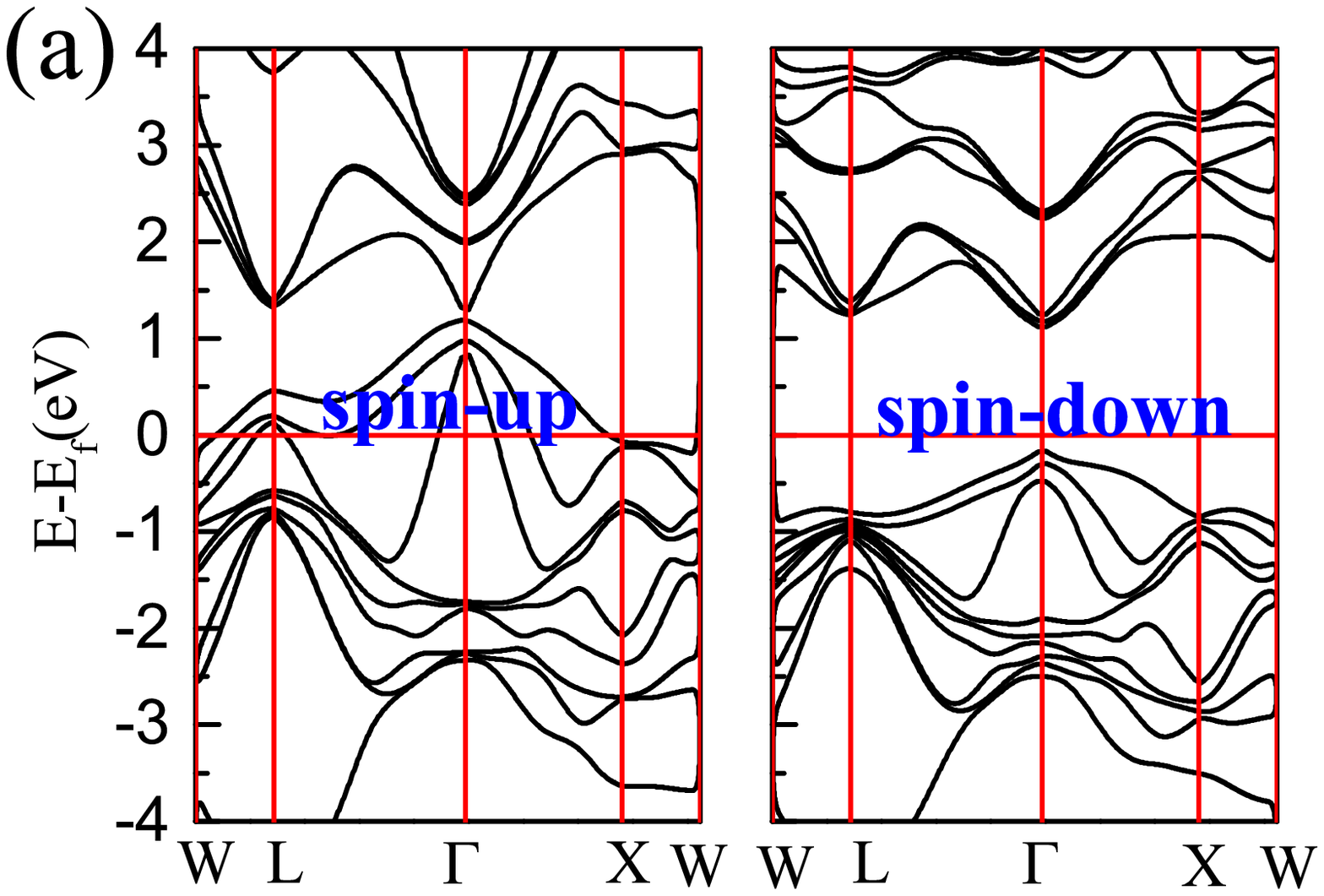}
  \includegraphics[width=.4\textwidth]{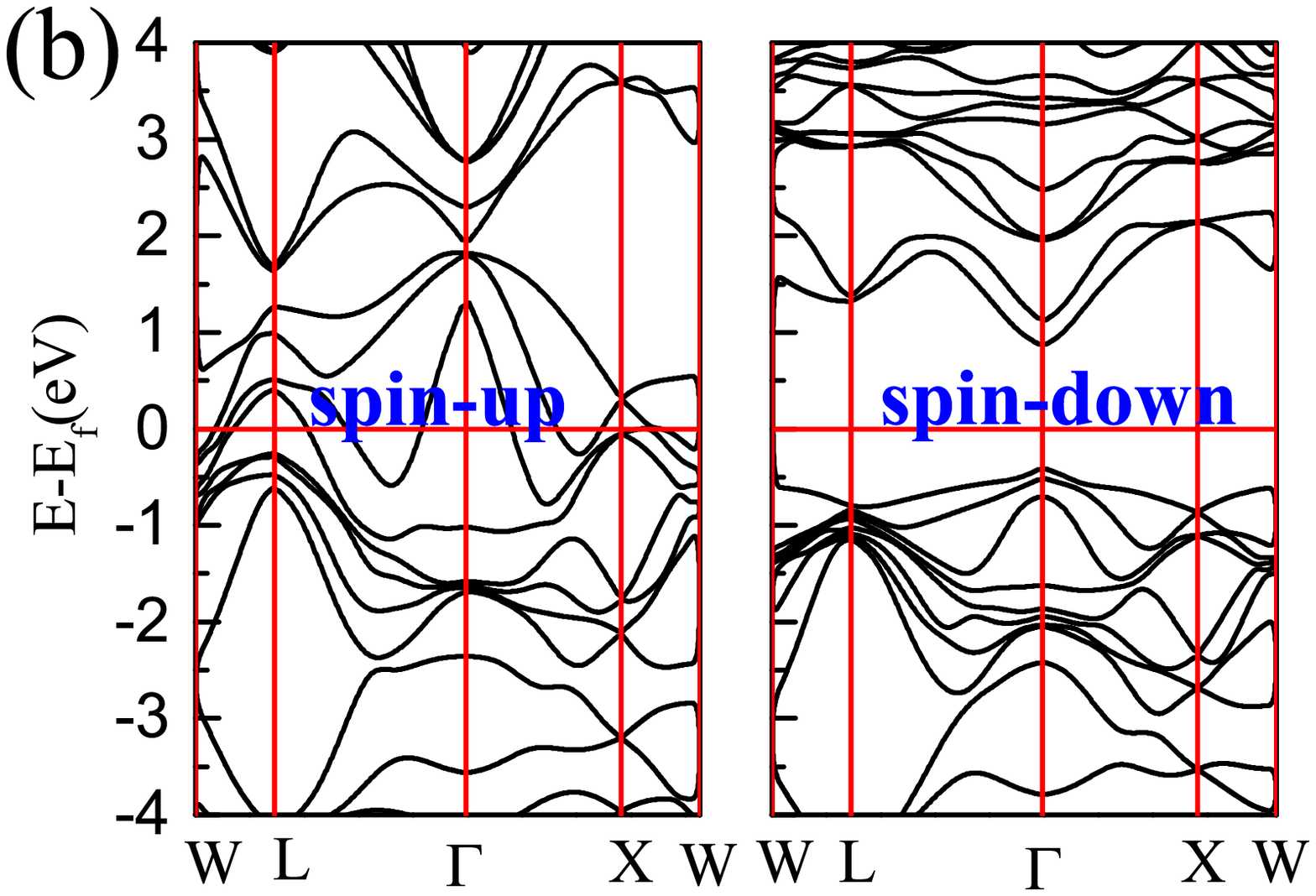}
  \includegraphics[width=.4\textwidth]{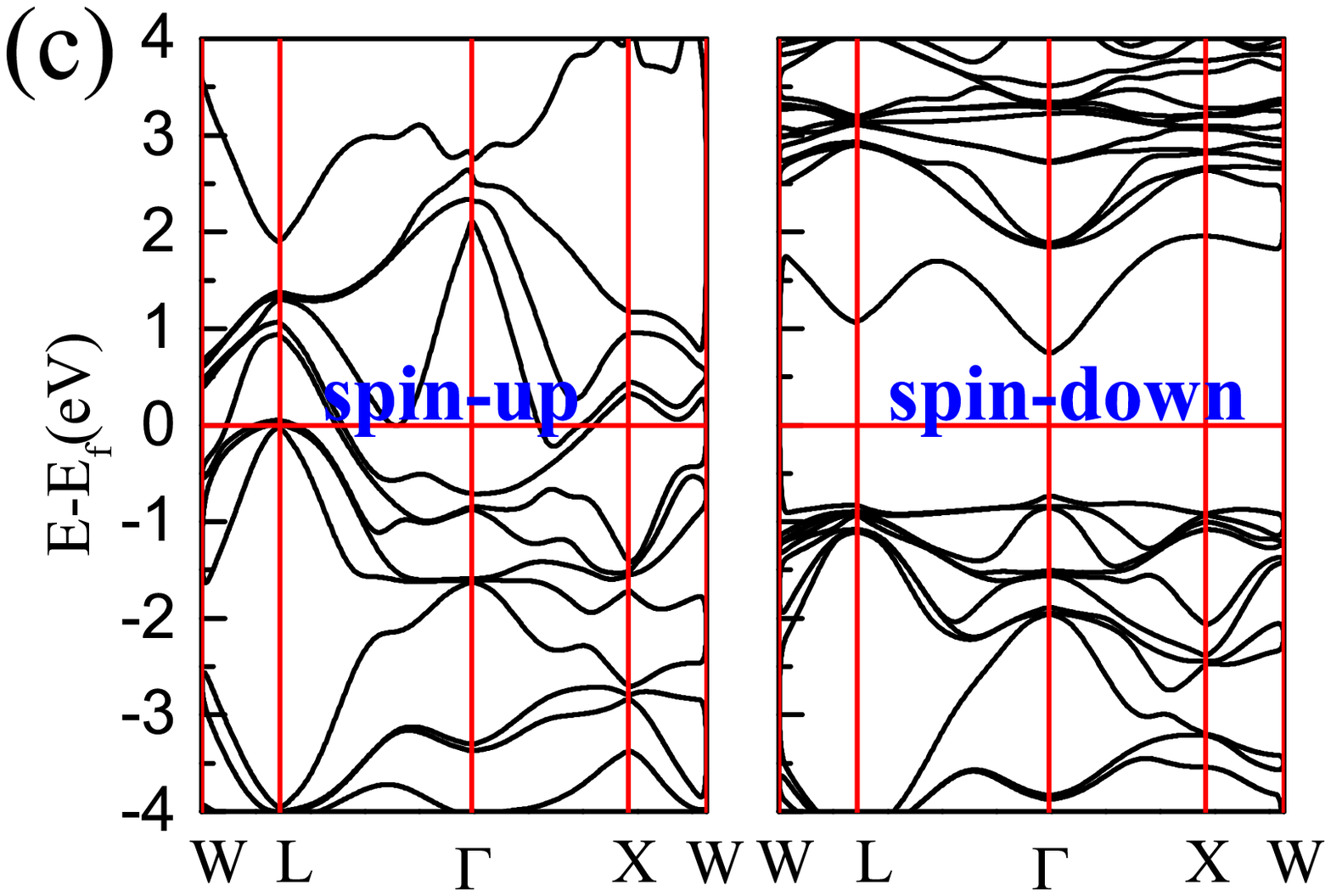}
  \includegraphics[width=.4\textwidth]{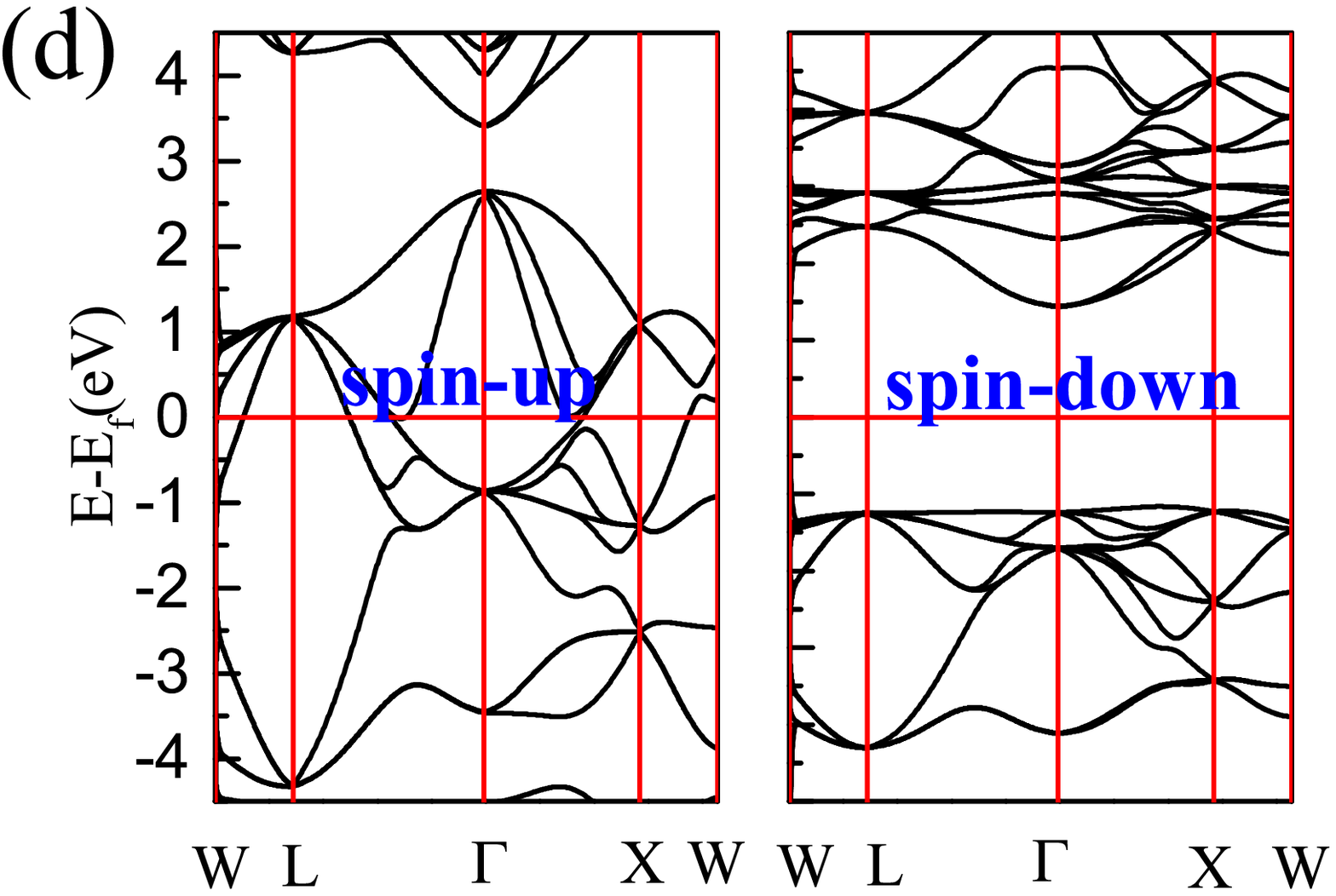}

  \caption{Spin-dependent band structures of (a) Ga$_{0.75}$Mn$_{0.25}$Sb; (b) Ga$_{0.75}$Mn$_{0.25}$Sb; (c) Ga$_{0.25}$Mn$_{0.75}$Sb; (d) MnSb.}\label{fig4}
\end{figure}

To further study the principle underlying the HM properties of this series of HMF materials, the total density of states of Ga$_{0.75}$Mn$_{0.25}$Sb is selected because the electronic densities of the Ga$_{1-x}$Mn$_{x}$Sb ($x = 0.25, 0.5, 0.75$) compounds are similar. Additionally, the atomic wave partial density of states is taken as an example. As Fig. 5 shows, the total electronic density of Ga$_{0.75}$Mn$_{0.25}$Sb is metallic in the spin state near the Fermi surface, while the lower spin state has a clear band gap, indicating semiconductivity. Since only one spin-oriented electron exists at the Fermi level of Ga$_{0.75}$Mn$_{0.25}$Sb, the spin polarizability is defined as~\cite{14,15}.

\begin{eqnarray}
P&=& \frac{N_{\uparrow}(E)-N_{\downarrow}(E)}{N_{\uparrow}(E)+N_{\downarrow}(E)}
 \label{equ2}
\end{eqnarray}

where  and  are the spin-up and spin-down density of states, respectively. We can conclude that the spin polarization of the conduction electrons of Ga$_{0.75}$Mn$_{0.25}$Sb is $100\%$.

Because the electromagnetic and optical properties of most materials are derived from the intermetallic orbital $pd$, $dd$, and $spd$ electron orbital hybridization, these properties are related to the electronic configuration of each atom and the electron orbital hybridization between atoms~\cite{52,53,54,55}. Because the valence electronic configuration of Mn is $3d^{5}4s^{2}$, and the valence electron configuration of Ga is $4s^{2}4p^{1}$, the outermost $4p$ orbit has only one electron in Ga. Additionally, the valence electron configuration of Sb is $5s^{2}5p^{3}$, and its $5p$ state has three electrons, which is the half-filled state ($p$-state full-shell layer is six electrons). Furthermore, comparing the bond length and density of states between individual atoms in the crystal suggests that strong hybridization occurs near the Fermi level. As Table II shows, the Mn-Sb bond length are shorter than that of the the Ga-Sb, indicating that the electron hybridization occurs mainly between the transition metal atom Mn and the Sb atom. The $3d$ orbital of Mn and the $5p$ orbital of Sb are found to undergo pd electron hybridization; therefore, the $3d$ electronic density distribution of Mn exhibits a shift in the energy level. In other words, electron orbital hybridization causes the total electronic density of crystal materials to redistribute, which is the main reason Ga$_{1-x}$Mn$_x$Sb ($x = 0.25, 0.5, 0.75$) exhibits FM HM properties.

\begin{figure}[htb]
  \centering
  \includegraphics[width=.5\textwidth]{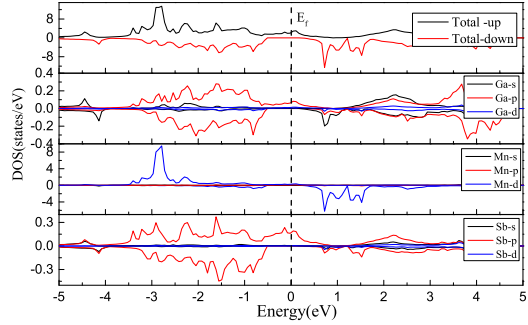}
  \caption{Total electron density states and projected density states of Ga$_{0.75}$Mn$_{0.25}$Sb.}\label{fig5}
\end{figure}

Optical Properties. The optical properties of a solid can be described by a dielectric function~\cite{56,57}, which consists of a real part and an imaginary part and is defined as

 \begin{eqnarray}
\varepsilon(\omega)&=&\varepsilon_{1}(\omega)+i\varepsilon_{2}(\omega)\label{equ3}
\end{eqnarray}

\begin{figure}[htb]
  \centering
  \includegraphics[width=.5\textwidth]{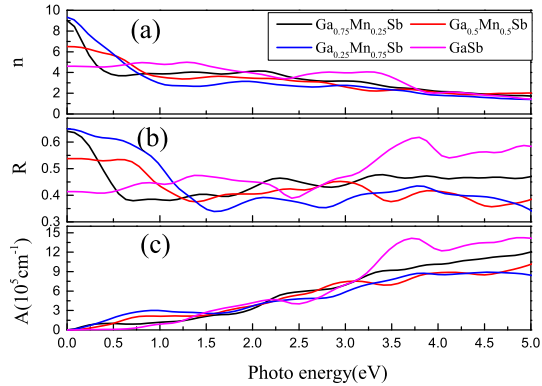}
  \caption{Comparison of the calculated optical properties of Ga$_{1-x}$Mn$_{x}$Sb ($x = 0, 0.25, 0.5, 0.75$): (a) optical refractive index; (a) optical reflectivity; (c) optical absorption coefficient.}\label{fig6}
\end{figure}

Because Ga$_{1-x}$Mn$_{x}$Sb ($x = 0.25, 0.5, 0.75$) belongs to the cubic system, its optical properties are known to be isotropic. In this study, the dielectric constant of Ga$_{1-x}$Mn$_{x}$Sb ($x = 0.25, 0.5, 0.75$) is calculated, and the optical properties are analyzed. Fig.6 shows the relationship between the optical properties and the incident light energy, comparing the refractive index n, reflectance R, and absorption coefficient A of some crystalline materials. We can use the real and imaginary parts of the dielectric function to obtain the absorption coefficient and reflectivity of the solid using the following formulas:

\begin{eqnarray}
\alpha(\omega)&=&\sqrt{2}\omega  \sqrt{\sqrt{\varepsilon_{1}^{2}(\omega)+\varepsilon_{2}^{2}(\omega)}-\varepsilon_{1}(\omega)}\label{equ4}
\end{eqnarray}

\begin{eqnarray}
R(\omega)&=& 	|\frac{\sqrt{\varepsilon(\omega)}-1}{\sqrt{\varepsilon(\omega)}+1}|^{2}  \label{equ5}
\end{eqnarray}

\begin{table}[h]
\begin{ruledtabular}
\caption{Crystal properties of Ga$_{1-x}$Mn$_{x}$Sb ,the equilibrium lattics constant a$_{0}$ ($\AA$), Mn-Sb bond length L$_{MS}$ ($\AA$), Ga-Sb bond length L$_{GS}$ ($\AA$), the HM gaps calculated by HSE and PBE, E$_{HM-HSE}$ (eV) and E$_{HM-PBE}$ (eV), respectively, the semiconductor gaps calculated by HSE and PBE, E$_{HSE}$ (eV) and E$_{PBE}$ (eV), respectively. }\label{tableII}
\begin{tabular}{lcccccccccc}
$x$ & a$_{0}$ & L$_{MS}$& L$_{GS}$ &E$_{HM-HSE}$ &E$_{HM-PBE}$& E$_{HSE}$& E$_{PBE}$\\
\hline
0    & 6.095   &  -      &2.554  &  -     &   -    &0.526   &0.083\\
     &         &         &       &        &        &0.720~\cite{58}&0.110~\cite{58}\\
0.25 & 6.208   & 2.663   &2.697  &  0.144 & 0.086  &1.239   &0.593\\
0.5  & 6.315   & 2.678   &2.689  &  0.405 &  -     &1.260   &0.452\\
0.75 & 6.179   & 2.671   &2.696  &  0.723 &  -     &1.455   &0.746\\
1    & 6.178   & 2.675   &  -    &  0.988 &  -     &2.217   &1.457\\
     & 6.166~\cite{59}  &   &  &  &  0.200~\cite{59} & &1.491~\cite{59}\\
     & 6.166~\cite{60}  &   &  &  &  0.150~\cite{60} & &0.900~\cite{60}\\
\end{tabular}
\end{ruledtabular}
\end{table}

Fig6(a) and 6(b) reveal that the static refractive index n(0)=9 of Ga$_{0.75}$Mn$_{0.25}$Sb and the static reflectance are R(0)=0.64, respectively, which are higher than those of GaSb, i.e., n(0)=4.6 and R(0)=0.41, respectively. The calculation results revealed that the static refractive index and static reflectivity of Mn-doped FM materials are higher than those of GaSb. The refraction spectrum of Ga$_{1-x}$Mn$_{x}$Sb ($x = 0.25, 0.5, 0.75$) in Fig. 6(a) demonstrates that in the energy range of $0-3.5$ eV, the refractive indices of these materials tend to decrease gradually with the increasing incident light energy, whereas they decrease rapidly in the energy range of visible light at $1.3-3.5$ eV. Furthermore, the refractive power of the material for infrared light is much higher than that for ultraviolet light, and the static refractive index and the static reflectivity of the doped material are both higher than those of undoped GaSb. The reflection and refraction patterns of different Mn concentrations demonstrate that the variations in the refractive index and reflectivity are similar when doping is not higher than $50\%$, and the higher the doping concentration is, the more obviously the refractive index changes.
Fig.6(c) reveals that in the energy range of $0-5$ eV, the absorption coefficients of the four Ga$_{1-x}$Mn$_{x}$Sb compounds tend to increase gradually. In the visible region, there is no obvious absorption peak, and the absorption coefficient of the material does not change much after doping with Mn. In contrast, in the infrared region, that is, when the incident photon energy below 1.3 eV, the effects of doping are obvious. The doped material absorbs significantly more infrared light than GaSb. Therefore, Ga$_{1-x}$Mn$_{x}$Sb ($x = 0.25, 0.5, 0.75$) has a strong absorption capacity for infrared light and is suitable for use in electronic devices related to infrared detection.

This paper presents the electronic structure and the magnetic and optical properties of Mn-doped semiconductor GaSb studied using a first-principles calculation method. The spin-polarized electronic density of states of Ga$_{1-x}$Mn$_{x}$Sb ($x = 0.25, 0.5, 0.75$) is calculated, and the electron band structure and magnetic moment show that Ga$_{1-x}$Mn$_{x}$Sb ($x = 0.25, 0.5, 0.75$) shows HM ferromagnetism, electron spin polarization of 100\% with an integer magnetic moment, a wide energy gap, and a HM gap. Additionally, the high Curie temperature indicates that HM Ga$_{1-x}$Mn$_{x}$Sb can be widely used in spintronic devices. Furthermore, optical properties such as the reflection and refraction of this series of materials were studied, and the optical properties of the HM materials Ga$_{1-x}$Mn$_{x}$Sb are found to be similar, with a large static reflection and static refractive index. In the visible light range, the absorption coefficient is small, but the infrared light absorption is strong. Therefore, the FM HM Ga$_{1-x}$Mn$_{x}$Sb series is also a potential material for infrared photoelectric devices.

\begin{acknowledgments}
 This work was supported by the NSFC (Grants No.11874273), the Specialized Research Fund for the Doctoral Program of Higher Education of China (Grant No.2018M631760), the Project of Hebei Educational Department, China (No. ZD2018015 and QN2018012), the Advanced Postdoctoral Programs of Hebei Province (No.B2017003004) and the Key Project of Sichuan Science and Technology Program (19YFSY0044). The numerical calculations in this paper have been done on the supercomputing system in the High Performance Computing Center of Yanshan University.
\end{acknowledgments}



\nocite{*}
\bibliography{wangchuang}
\end{document}